\documentclass[aps,prab,twocolumn,superscriptaddress]{revtex4-1}

\usepackage{amssymb}
\usepackage{lipsum}
\usepackage{amssymb}
\usepackage{bbold}
\usepackage{amssymb}
\usepackage{graphicx}
\usepackage{graphics}
\usepackage{amsmath}
\usepackage{placeins}
\usepackage{hyperref}
\usepackage[dvipsnames]{xcolor}
\usepackage{wasysym}
\usepackage{soul}
\usepackage{float}
\usepackage{array}
\usepackage{tabulary}
\usepackage{caption}
\usepackage{fancyhdr}
\usepackage{wrapfig}

\usepackage[english,ngerman]{babel}

\usepackage[normalem]{ulem}
\usepackage{graphicx}
\usepackage{color,soul}
\usepackage{subfigure}
\usepackage{dcolumn}
\usepackage{bm}
\usepackage{hyperref}
\usepackage{url}
\usepackage{multirow}
\hypersetup{
    colorlinks=true,
    linkcolor=blue,
    filecolor=blue,      
    urlcolor=blue,
    citecolor=blue,
}

\usepackage[makeroom]{cancel}

\begin{document}
\selectlanguage{english}
\title{Thermal Emittance Isolation by Cathode Retraction}

    \author{\firstname{Benjamin} \surname{Sims}}
    \email{simsben1@msu.edu}
    \affiliation{Department of Electrical and Computer Engineering, Michigan State University, MI 48824, USA}
    \affiliation{Department of Physics and Astronomy, Michigan State University, East Lansing, MI 48824, USA}
    \affiliation{Facility for Rare Isotope Beams, Michigan State University, East Lansing, MI 48824, USA}

    \author{\firstname{John W.} \surname{Lewellen}}
    \email{jwlewellen@lanl.gov}
    \affiliation{Accelerator Operations and Technology Division, Los Alamos National Laboratory, NM 87545, USA}
    
    \author{\firstname{Xu} \surname{Ting}}
    \email{XuTi@frib.msu.edu}
    \affiliation{Department of Physics and Astronomy, Michigan State University, East Lansing, MI 48824, USA}
    \affiliation{Facility for Rare Isotope Beams, Michigan State University, East Lansing, MI 48824, USA}
    
    \author{\firstname{Sergey V.} \surname{Baryshev}}
    \email{serbar@msu.edu}
	\affiliation{Department of Electrical and Computer Engineering, Michigan State University, MI 48824, USA}
	\affiliation{Department of Chemical Engineering and Material Science, Michigan State University, MI 48824, USA}

\begin{abstract}

    In this work, a combination of cathode retraction and two-slit emittance measurement technique is proposed as an advanced means to individually modify emittance growth components, specifically, rf injector fringe fields, to isolate and directly measure the thermal emittance, the fundamental beam emittance metric for an electron beam. A case study of the LCLS-II-HE Low Emittance Injector (LEI), a state-of-the-art superconducting radiofrequency (SRF) gun, designed for LCLS-II HE upgrade is used to showcase the power of the two-slit technique. Particularly, it is demonstrated that generating a high resolution phase-space distribution map, dominated by the intrinsic emittance of the electron bunch, is possible. This result goes beyond the normal single-parameter distribution characterizations (e.g. RMS emittance and Twiss parameters) provided by the solenoid scan.
    One key feature making this technique work (and in the end practically useful) is the ability to retract the cathode, because it provides the ability to compensate for radiofrequency (rf) de-focusing. It is demonstrated how the cathode retraction can serve as an additional optimisation tool for tailoring the routine performance of the photoinjector. We posit that a variable position cathode may be a useful method for optimizing photoinjector performance across multiple parameters regimes.

\end{abstract}

\maketitle

\section{Introduction}\label{intro}
Superconducting radiofrequency (SRF) injectors are emerging as powerful sources for applications requiring high-average-current and high-brightness electron beams.  Use cases include tools of scientific discovery (e.g.  as beam sources for ultrafast electron diffraction and microscopy systems, and X-ray free electron laser injectors), as well as medical and industrial applications requiring high duty factors and average beam power \cite{nosochkov:ipac2022-tupopt046, PhysRevAccelBeams.22.023403, PhysRevSTAB.17.104401,PhysRevSTAB.15.090701}. While several metrics, such as beam brightness, have been established to characterize the quality of electron beams used for scientific applications, a parameter termed emittance is a central figure of merit for many applications. Roughly analogous to the M-squared figure-of-merit used for laser beams, the transverse emittance of an electron beam provides a characterization of the transverse phase-space density of the beam.  An ideal electron beam would have an emittance limited by the Fermi exclusion principle.  In practice, the emittance of a beam is determined by a number of factors, including contributions from space charge forces within the beam, nonlinear radiofrequency (rf) and static electromagnetic fields with which the beam interacts, and intrinsic emittance arising from the conditions during the beam's initial emission from a solid-state cathode \cite{sims_emittance_2023}.

Because in legacy systems beam emittance growth has been dominated by the space charge and nonlinear fields,  efficient methods have been found to manage and minimize (or compensate) them \cite{bazarov_benchmarking_2008}. Current beam source designs are approaching the point where intrinsic emittance can become a limiting factor in beam quality. Since the intrinsic emittance depends on the cathode material used, it needs to be measured experimentally as a benchmark of the emittance floor, i.e. setting the best attainable case. Techniques and apparatus such as the momentatron \cite{10.1063/5.0013122} have been developed to do this under "laboratory" conditions. However, experimentally measuring the intrinsic emittance in an operational beam source can be difficult due to the contributions from space charge and nonlinear external fields \cite{ma_transverse_2022}.  Given the relatively fragile nature of high-quantum-efficiency photocathodes, the ability to characterize in detail the evolution of cathode performance over time in an operational setting is increasingly essential.

The solenoid scan technique \cite{anderson_space-charge_2002, hachmann_transverse_2012, pinayev_solenoid_nodate} has been a workhorse for measuring emittance because solenoid scans are fast, most if not all of the equipment is already present in most high-brightness photoinjector beamlines, and the technique itself is straightforward to implement. However, measuring emittance using solenoids has limitations associated with aberrations inherent to the solenoidal field, typically leading to potentially significant intrinsic emittance overestimation \cite{zheng_overestimation_2018, zheng_eliminating_2022, ma_transverse_2022,}. As usually implemented, solenoid scans provide only a single $cumulative$ (or whole-bunch) emittance measurement, i.e. without providing information about the detailed transverse phase-space distribution \cite{lu_transverse_2013}.  To extract RMS (whole-beam) parameters, additional assumptions may be required as the solenoid scan technique alters the impact of space charge during the measurement process, most apparently when the beam is focused to a waist, giving rise to uncertainties when estimating the beam parameters \cite{anderson_space-charge_2002}. For very low emittance beams, resolution of the beam imaging system employed may also prove to be a limiting factor. 

Such issues aside, a density map of the beam's transverse phase space would be a preferable measurement as it could contain important details about individual emittance contributions, as well as cathode emission characteristics, that could lead to better informed injector R\&D and performance tuning. Tomographic techniques \cite{yakimenko_electron_2003, stratakis_tomographic_2007, xiang_transverse_2009} have been used to develop phase-space maps, but these techniques have generally been applied at higher beam energies, such that beam has evolved significantly since its initial formation.  While the starting phase-space distribution can be attributed to the cathode properties (e.g. intrinsic emittance and quantum efficiency) convolved with the drive laser, emittance growth over time can be attributed to space charge, acceleration and propagation through nonlinear fields, etc. Tomographic measurements performed at the photoinjector using a solenoid are also still subject to the effects mentioned above, e.g. spherical aberration.
In this context, an arguably better way to measure the intrinsic emittance immediately after a photoinjector is by using the two-slit technique \cite{anderson_space-charge_2002, ma_application_2023}. Given the phase space maps can be directly obtained, it opens up an opportunity to quantify additional lingering effects of space charge and rf emittance components that can be additionally manipulated by an effective and practical means such as cathode retraction with respect to the injector back wall \cite{Herrmannsfeldt:1989fi}. As the goal is to measure the intrinsic emittance, ideally to obtain a two-dimensional phase-space density map rather than a single RMS value, a slit method is preferable as long as it can provide the required resolution within a feasible measurement time. Contrary to the rapid solenoid scan, two-slit measurement comes at a cost in terms of time, where the duration of the measurement and errors resulting from drifts and jitter are major, and coupled, liming factors \cite{wang_design_2019}. Indeed, reducing the intrinsic measurement error requires higher resolution scans that, consequently, increases the duration of the scan. Results from such long duration scans could be subject to system drifts or instabilities like limited lifetime of high quantum efficiency (QE) photocathodes. Nevertheless, the detailed phase-space distribution provides a substantially improved ability to identify unexpected behaviors via direct inspection of the phase-space distribution, in addition to providing the ability to calculate the emittance from the measured distribution. Therefore, finding optimal ways to perform two-slit measurements deserves further efforts.

In the present work, using an example of the low emittance SRF injector being developed for LCLS-II HE upgrade \cite{nosochkov:ipac2022-tupopt046}, we computationally demonstrate a novel approach to intrinsic emittance measurement that utilizes $i)$ cathode retraction to compensate for rf-induced emittance growth, and $ii)$ a 2-slit emittance measurement system to obtain a phase-space map. Practically speaking, the proposed diagnostic two-slit beamline could serve to characterize high brightness bunches and provide a means to characterize high-brightness photocathodes in an operating photoinjector.

\section{Definitions of Emittance and Error}

\subsection{Two Slit Emittance Measurement}\label{2 Slit}
A two slit emittance measurement works by measuring the intensity of "bunchlets," located at position $x$ and angle $x'$ with widths $\delta x$ and $\delta x'$ respectively within the beam bunch; the ensemble of measurements form a current density map of the transverse phase space $\rho (x,x')$. (Strictly, we are measuring a projection of the full 6-d phase space distribution of the bunch $\rho (x,x',y,y',p,t)$, onto a single plane.) The apparatus consists of two plates with transverse slits (simply referred to as "slits" hereafter) located along the axis of beam propagation.  A detector is located downstream of the second slit; the detector can be an optical screen, a Faraday cup, etc. The slit plates are made thick enough to absorb beam particles not entering the slit, but thin enough so as not to significantly collimate the beam.

The transverse position of the first, or upstream, slit $X_1$ sets the position of the bunchlet center, that is $x$=$X_1$, and the difference in the center positions of the upstream and downstream slits, $X_1$ and $X_2$ respectively,  divided by the distance between the slits, $L$, determines the angle $x'$ as \cite{zhang_emittance_1996, lewellen_electrostatic_2018}
\begin{gather}
    x'= \frac{X_2 - X_1}{L} \label{x'}
\end{gather}
The number of transmitted beam particles at each pair of slit locations $X_1$ and $X_2$ (or the corresponding location in the beam's phase space  $x$ and $x'$) describes a data bin. These bins can be combined together into a density map representing the bunch phase space $\rho (x,x')$. The resolution of the phase space map is determined by the slit widths ($W_1$ and $W_2$) and longitudinal separation of the slits $L$ as
 \begin{gather}
    \Delta x= W_1 \label{x_res} \\
    \Delta x'= \frac{W_2}{L} \label{x'_res}
    \end{gather}
This indicates that small slits and large separation $L$ between slits will provide higher resolution, e.g. more bins across a given phase-space distribution.  The bin size is analogous to pixel size in a conventional imaging system. 

\label{idea}

Two-slit measurements of the intrinsic emittance require the management of space charge and rf induced emittance growth, such that the emittance of the beam is dominated by the intrinsic emittance. Space charge-induced growth can be minimized by using low-intensity laser pulses, i.e., by conducting measurements at low bunch charge.
Additionally, by sweeping laser parameters, specifically the intensity, we can assess the impact of space charge, effectively characterizing its contribution to the overall emittance.

The intrinsic emittance can be increased by increasing the primary laser spot size, making the intrinsic emittance easier to measure. This, however, is not without a cost as the increase in radius also increases the emittance contribution from nonlinear fields, as well as (all other effects equal) increasing the size of the beam spot at the first slit. This effect can be addressed with the use of cathode retraction. By retracting the cathode, a transverse focusing field near the cathode is introduced by the aperture edges. Such lensing effects helps mitigate the rf defocusing of the beam as it exits the rf gun, allowing for the use of a larger laser spot on the cathode. There should exist an optimal location where focusing provided by cathode retraction and rf defocusing effects from the rf field, in the body and at the exit of the gun, effectively cancel each other, providing a net minimization of beam divergence due to rf effects. Hence, rf emittance contribution to the total emittance is minimized. This approach also results in a smaller spot at the first slit, with a smaller far-field divergence angle, which is beneficial for the two-slit method.  We note that while a solenoid located immediately downstream of the gun cavity, e.g. in the typical location for emittance compensation, can provide a small beam at the first slit, it cannot provide the same benefits in terms of minimizing nonlinear field contributions as the method of cathode retraction can.

\subsection{Binning Error}
$Edge$ $bins$, those located near the edges of the beam in phase-space, represent a source of error similar to that encountered in particle-in-cell space-charge calculations.
All particles within a bin are typically assigned to a single ($x$, $x'$) determined by the slit locations. For reasonable bin sizes, bins with many particles, and beams with relatively slow variations across the phase-space distribution, binning is a reasonable approach for phase space mapping. However, edge bins near the boundaries of the distribution may collect a relatively small number of particles (or even none in low bunch charge case), and the center of the bin may not correspond well to the actual average position of the particles within the bin. Thus, the error of the two-slit measurement can be directly correlated to the number of edge bins as \cite{ludwig_quantization_1994} where the error is proportional to both the resolution and the maximum $x$ and $x'$ measured. The result of the difference between these two values, as seen in Eq.\ref{Error}, is the approximate emittance without the error attributable to binning effects.

 \begin{gather}
   \varepsilon_{er} = \frac{n_{edge} \cdot \Delta x \cdot \Delta x' }{2 \cdot \pi}. \label{Error_nb}  
\end{gather}
The emittance measured by any two-slit scan can be approximated by \cite{ludwig_quantization_1994}
 \begin{gather}
    \varepsilon = \frac{n \cdot \Delta x \cdot \Delta x' }{\pi}. \label{MeasuredEmittance}      
\end{gather}
When combined with additional metric for calculating the error in the measured emittance, it yields
 \begin{gather}
    \varepsilon_{er} = \frac{2}{\pi} \cdot (x_{max} \cdot \Delta x' - x'_{max} \cdot \Delta x ), \label{Error}     
\end{gather}

The set of the given equations establishes the basis for informed design of experimental beamline where slit-based phase space measurements with minimized error.

\subsection{Basic Emittance Concepts}\label{Emittance}
Projected transverse emittance can be calculated using either velocity or momentum phase space. As many of the phase spaces measured are non-elliptical, momentum space was chosen to help account for any irregularities that would be overlooked in the velocity space \cite{gpt}. The conversion from the measured velocity space to momentum space is done by multiplying the measured $x'$ by the Lorentz \(\gamma\)-factor of the particle. The statistical root mean square (RMS) and 100\% of the particles were used for emittance calculation as  
 \begin{gather}
    S_{11}= \overline{x \cdot x},\label{s11} \\
    S_{12}= \overline{x \cdot x'},\label{s12} \\
    S_{22}= \overline{x' \cdot x'},\label{s22} \\
    \varepsilon = \sqrt{S_{11} \cdot S_{22} - S_{12}^2  }. \label{emittance} 
\end{gather}
The unnormalized emittance (scales with the beam's \(\gamma\)-factor) versus the normalized emittance (e.g. nominally invariant under acceleration) were used, as it allows for direct comparison between the calculated emittance and that expected given the parameters of the cathode.

\subsection{Intrinsic Emittance and MTE}
The intrinsic emittance of a bunch is determined by the initial spot size of the bunch and the mean transverse energy (MTE) of electrons emitted from the cathode. The MTE is determined by several factors including the photocathode material and illumination wavelength, temperature, and surface roughness. In the presented simulations, the cathode MTE is chosen in accord with the requirements of the LCLS-II-HE low-emittance injector \cite{osti_1029479}.  
The intrinsic emittance of the beam can be calculated as:
 \begin{gather}
    \varepsilon_{int}=\sigma_{xi}\sqrt{\frac{2 \cdot MTE}{m_e c^2} },\label{e_int} \\
    \sigma_{xi} = \frac{R_i}{2}, \label{R} \\
    \varepsilon_{int}=\frac{R_i}{2}\sqrt{\frac{2 \cdot MTE}{m_e c^2} },\label{e_int_r} \\
    MTE= \frac{m_e \cdot c^2 }{2} \cdot  \bigg({\frac{2 \cdot \varepsilon_{int}}{R_i}}\bigg)^2 ,\label{MTE} 
\end{gather}
where, assuming a uniform emission current density, $R_i$ is the emission spot radius. These equations allow for the calculation of the expected intrinsic emittance for a given MTE and emission spot radius, and thus the contribution of the MTE to the total emittance and therefore can be compared against the simulated slit measurement.

\section{Case Study: Low Emittance Injector for LCLS-II-HE upgrade}\label{LEI}

\subsection{Case study setup}\label{setup}
\begin{figure*}
	\includegraphics[width=7cm]{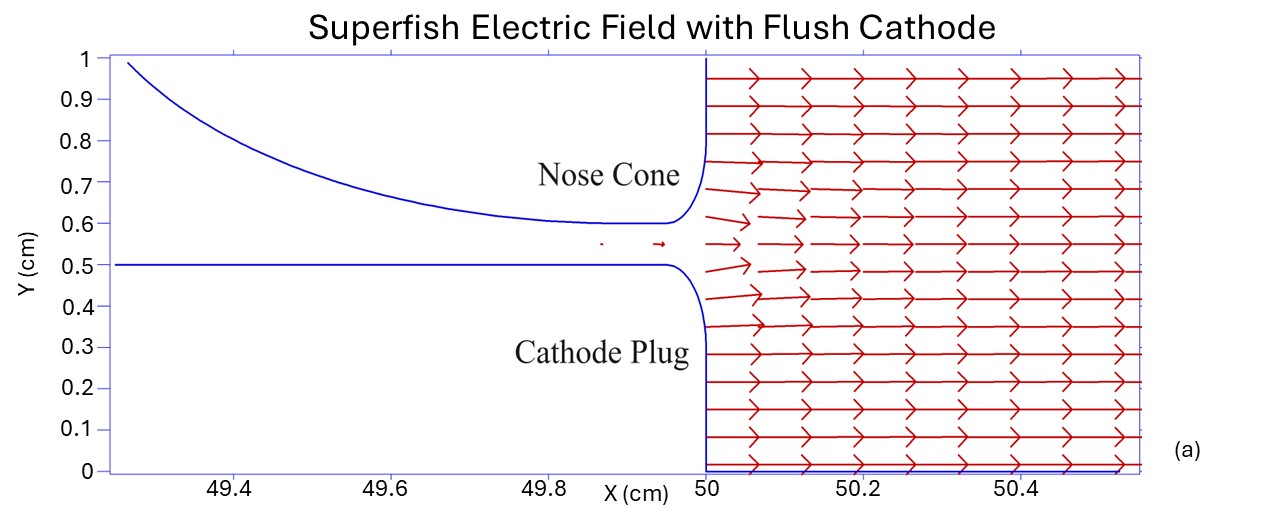}
	\includegraphics[width=7cm]{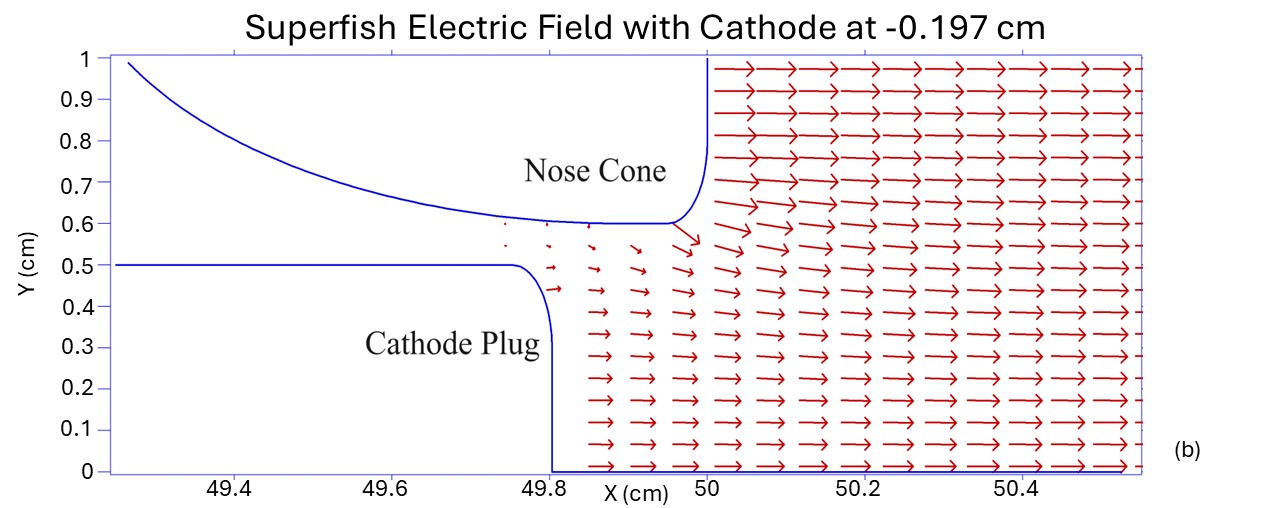}
	\caption{Superfish model of the LEI SRF gun with the cathode retracted (a) 0 cm, or flush with the nosecone face (the nominal operating position) and (b) retracted by -0.2 cm.}
	\label{LEI_-0.197}
\end{figure*}

The LCLS-II-HE low emittance injector (LEI) is a state-of-the-art high-gradient SRF injector design \cite{osti_1029479}. The LEI is intended to enable extending the LCLS-II-HE’s useful photon energy to 20 keV without additional cryomodules (e.g. increasing the beam energy past the LCLS-II-HE goal of 8 GeV) \cite{osti_1029479} by providing a significantly lower-emittance beam at 100 MeV, than the current LCLS-II injector. The LEI begins with a 1.8-MeV SRF photoinjector, being developed by a collaboration between SLAC, MSU/FRIB, Argonne and HZDR.  Low emittance bunch production necessitates this case study focusing on setting a useful and versatile photocathode testing beamline for the LEI. A robust two-slit emittance measurement optimized for the LEI SRF gun was considered. Requirements for any such system under consideration include compatibility and integrability with the current LEI gun-to-linac beamline design, and the ability to measure photocathode MTEs below 200 meV (e.g. suitable for cathodes proposed for the LEI ) \cite{osti_1029479}. $In~situ$ measurement of photocathode MTE, and evolution thereof, would then help attain best overall performance of the LEI. The design of the SRF gun allows for manipulation of the cathode stalk, in particular variation of the longitudinal position of the cathode surface relative to the gun "nosecone" surface \cite{osti_1029479}. Utilization of this feature is key to the MTE measurement process, as proposed here: the cathode-region fields, as modified by shifting the cathode longitudinal position, enable a low-error measurement.

The work was preformed utilizing Superfish \cite{SUPERFISH}, General Particle Tracer (GPT) \cite{GeneralParticelTracer}, and a proprietary sequencer. Superfish was used to generate rf field maps of the LEI SRF gun with the cathode surface located at different longitudinal positions. Several examples can be seen in Fig.\ref{LEI_-0.197}. The field maps were imported into GPT to simulate, visualize and quantify the two slit measurement. The sequencer was used to automate transition from Superfish to GPT and for quantifiable data collection and post-processing.

\subsection{Cathode retraction}

\begin{figure}
	\includegraphics[width=7cm]{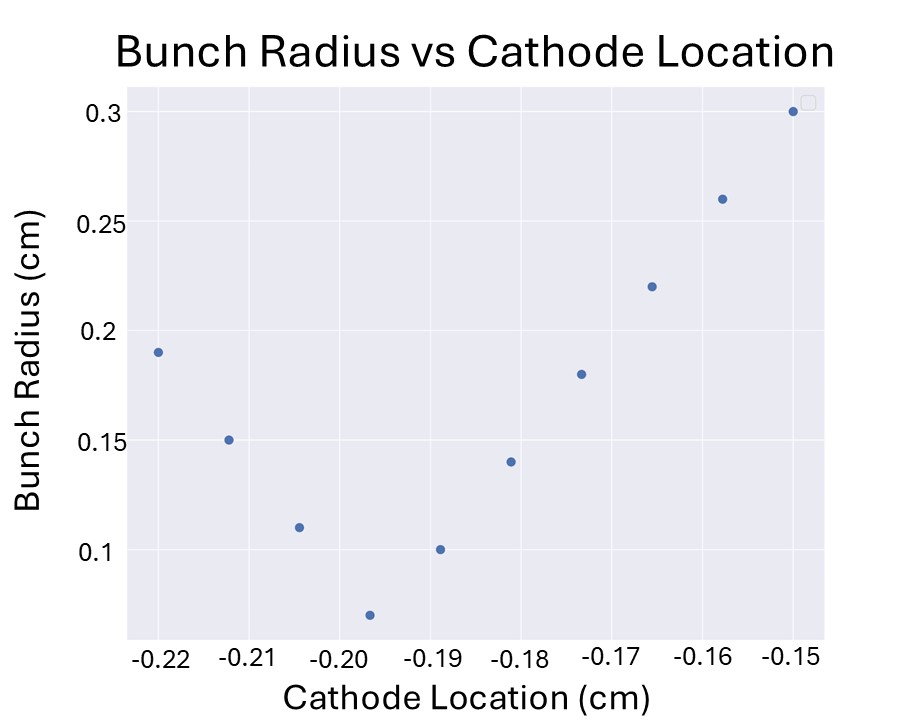}
	\caption{Simulated RMS bunch radius at 1 m for various cathode positions from the nominal cathode position (flush with nosecone).}
	\label{Bunch_Radius_vs_Cathode_Location}
\end{figure}

\begin{figure*}
	\includegraphics[width=7cm]{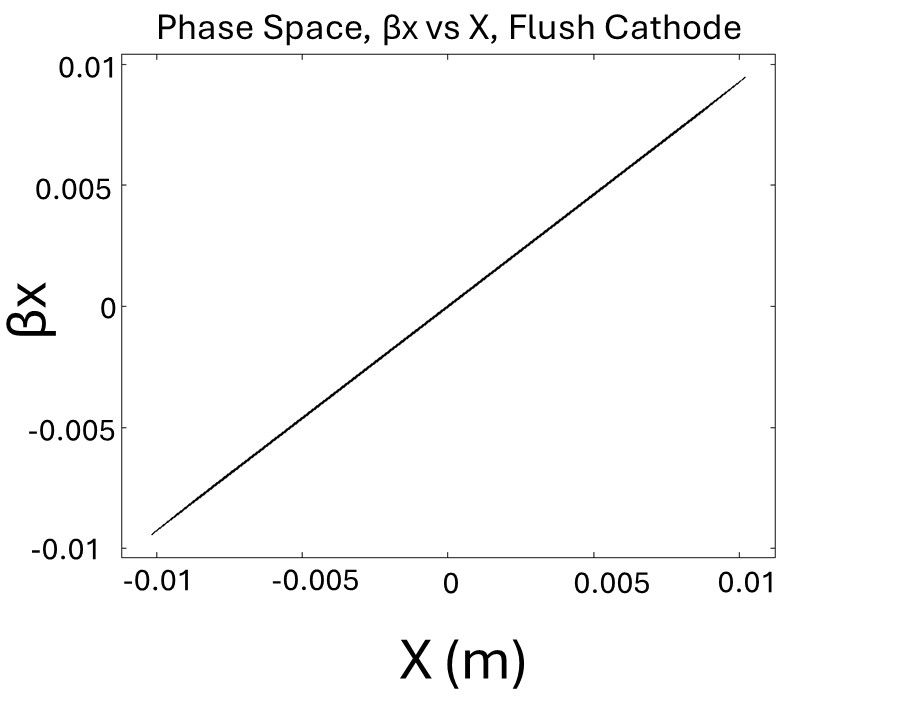}
	\includegraphics[width=7cm]{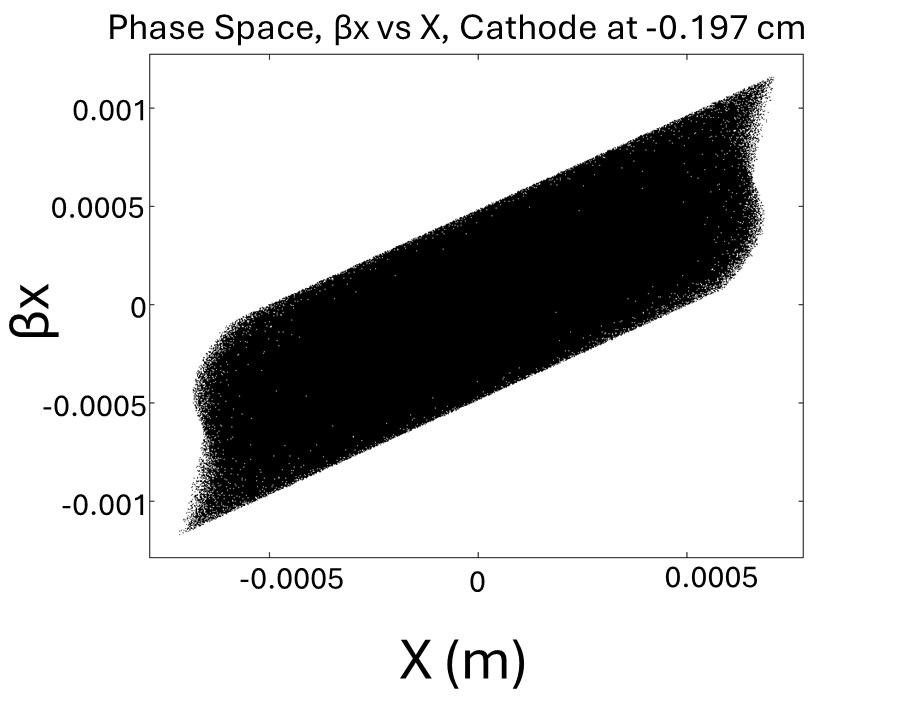}
	\caption{Bunch phase space with (a) cathode in the nominal position (e.g. no retraction, flush with nosecone) and (b) cathode retracted to --0.197 cm.} 
\label{Phase_Space_0.197}
\end{figure*}

Under nominal operating conditions for the LEI (e.g. 100-pC bunch charge), the SRF gun cathode is flush with the gun's nosecone surface.  This configuration produces a diverging beam, which is compensated by the gun solenoid as part of the traditional emittance-compensation process. However, for the proposed measurement technique to compensate for the divergence of the bunch, cathode retraction is used along with reduced-charge bunches.
This case study was conducted with the cathode being retracted from --0.15 cm to --0.22 cm with respect its flush (0~cm retraction) reference position, Fig.~\ref{Bunch_Radius_vs_Cathode_Location}. In this range, a minima was identified where the smallest bunch transverse size was observed on a screen located 1 m downstream from the injector exit plane; we define the corresponding cathode location as the optimal cathode retraction for intrinsic emittance measurement. The variation in spot size with cathode position is due to the emergence of the radial focusing fields that can be seen in Fig.~\ref{LEI_-0.197}. In accord with Eq.~\ref{Error}, the minimum spot size, corresponding to retraction of --0.197 mm is a desirable condition for measuring the intrinsic beam emittance because systemic error can be minimized. Fig.~\ref{Phase_Space_0.197} contrasts two bunch transverse phase spaces corresponding to a cathode retraction of zero, and the optimal position for intrinsic emittance measurement. A bunch emitted with the cathode in its nominal (flush) location is strongly diverging, with a "narrow" phase space as shown in Fig.\ref{Phase_Space_0.197}a. Attempting to measure this distribution produces a larger error in accordance with Eqs.\ref{Error}~and~\ref{Error_nb} as it has a large $x$ and $x'$ spread in addition to a high number of edge bins.
Alternately when the cathode is retracted the phase space at the screen is quite different, Fig.\ref{Phase_Space_0.197}. Retracting the cathode has two benefits when measuring MTE. First, it is seen that the extent of the phase space with the cathode at --0.197 cm is about an order of magnitude smaller in both $x$ and $x'$ than with the cathode in the nominal position. While the actual phase-space area in both cases is nominally the same, the potential measurement resolution is better and the number of edge bin is smaller, allowing for more accurate measurements. Second, the dominant factor in the bunch's transverse phase space is no longer the radial defocusing term from the gun's rf field. Instead the divergence is dominated by the MTE with vanishing contribution from the rf fields.  Note that, while we describe the beam location as at a "focus," e.g. smallest obtainable spot size, as a matter of convenience, the beam is not in fact at a waist but is diverging.  Thus, it meets one of the criteria for the 2-slit measurement technique, e.g. that the location of the first slit be downstream of a beam waist.

\subsection{GPT beamline modeling}\label{GPT}

\begin{figure}
	\includegraphics[width=7cm]{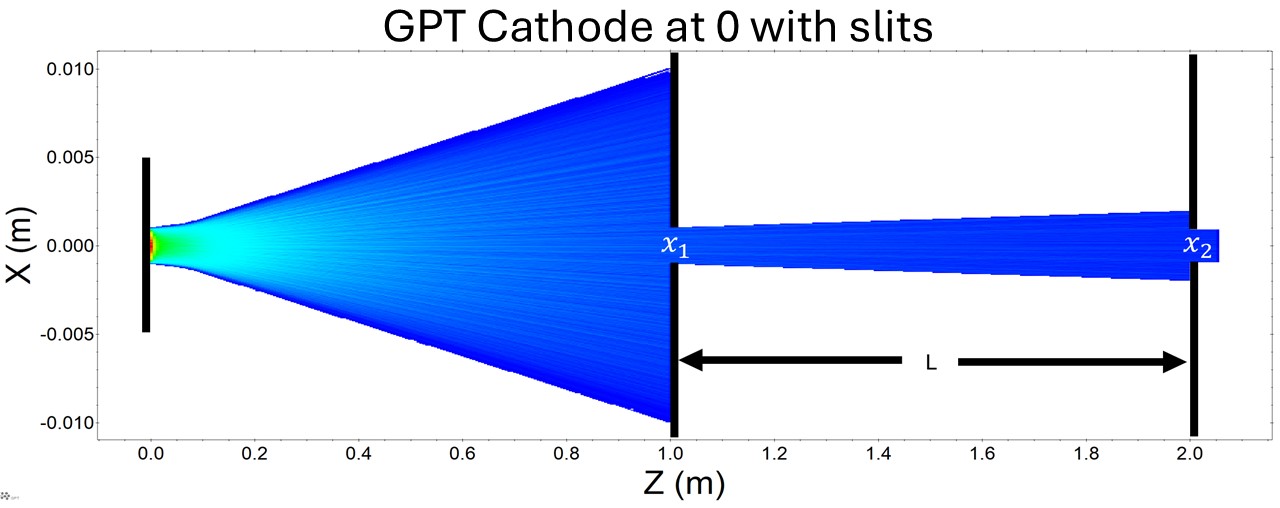}
	\caption{GPT simulation illustrating the effect of two exaggerated slits when the standard position is used, i.e. when cathode is flush with the gun's nosecone.}
	\label{GPT_SLITS}
\end{figure}

GPT simulations were performed with slits located at $z$=1 m and $z$=2 m, for a separation $L$=1 m. This set-up (Fig.~\ref{GPT_SLITS}) is sufficient to allow the beamlet passed by the first slit, to diverge appropriately. The largest spot size at each screen determined slit widths for all simulations as the parameters were, in analogy to a physical measurement with fixed-width slits, not modified over the course of the simulated measurement. The up- and down-stream slits were sized as $W_1$=99 \(\mu\)m and $W_2$=198 \(\mu\)m respectively, to provide 101 bins across the beam spot at both longitudinal locations. This was done as to accommodate the spot size produced with the cathode retracted to --0.15cm resulting in a spot size at $z$=1~m~of~0.5~cm and the spot size at $z$=2~m~of~1 cm. As the bunch is larger at the second screen, a larger slit size is needed to fully map it in the same number of steps as at first slit location. In an experimental setting, such slits can be readily fabricated.  With a nominal beam energy of 1.8 MeV, if made of tungsten the slits would need to be at least 1mm thick to completely stop the beam.  The angular acceptance of the first slit would therefore be approximately atan(0.1 mm / 1 mm) = 5.7 deg, and approximately 11.3 deg. for the second slit.  The angular resolution provided by the second slit is (0.2 mm / 1 m) = 0.01 deg; and the anticipated divergence of the beam as a whole, when the cathode is in the retracted location, is on the order of 1 deg.  Thus, we would not expect the angular resolution of the measurement to be limited by the upstream slit acting as a collimator.

In GPT, the bunch was generated as a uniform spot with a radius of 1~mm and a Gaussian temporal profile with a \(\sigma\) of 5.67 ps, clipped at $\pm 3 \sigma$. The MTE was set to 200 meV. Space-charge calculations were not included in the simulation, as a low-intensity laser pulse yielding a bunch charge of around 1 pC would have negligible space charge contributions to the emittance, while providing sufficient charge to be captured by a Faraday cup like device \cite{10.1063/5.0013172,8260552}, but space-charge calculation would significantly increase the time required to perform the simulations. Particles transmitted through both slits were counted at each pair of slit locations. This process was repeated for 10 different cathode retracted positions, between --0.15 cm to --0.22 cm, as seen in Fig.~\ref{Bunch_Radius_vs_Cathode_Location}.

Both the cathode-region focusing and exit-region defocusing are manifestations of the same phenomenon, e.g. a spatially varying longitudinal field gives rise to radial fields.  A longitudinal field increasing in magnitude (as in the case in the near-cathode region when the cathode is retracted) gives rise to a radially focusing field; while a longitudinal field decreasing in magnitude (as is the case along the majority of the axis within the gun) leads to a radially defocusing field \cite{griffiths2013introduction}.

\begin{figure}
	\includegraphics[width=7cm]{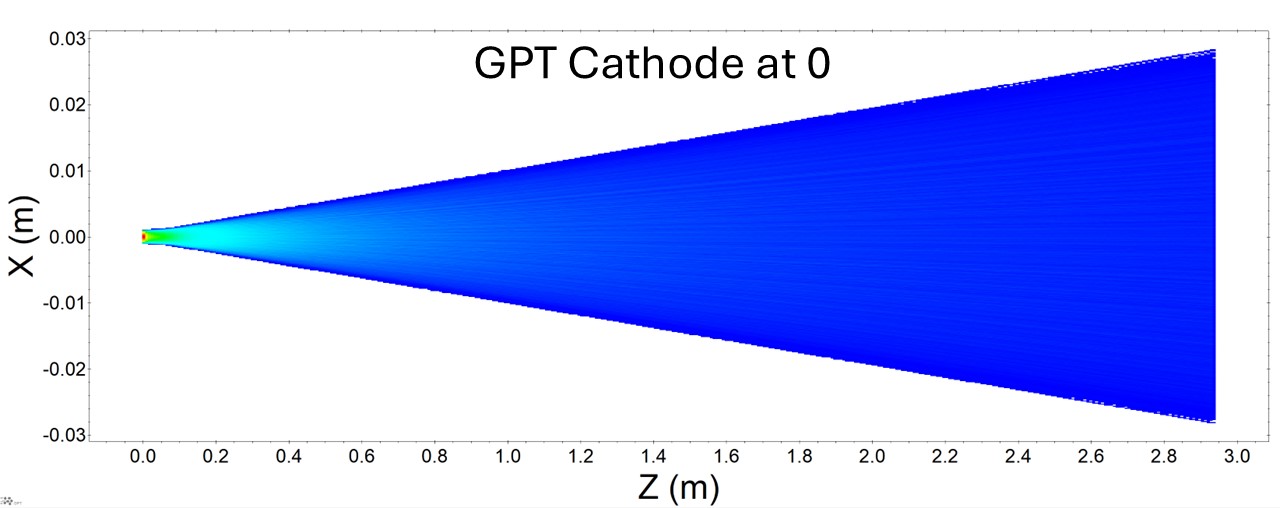}
	\includegraphics[width=7cm]{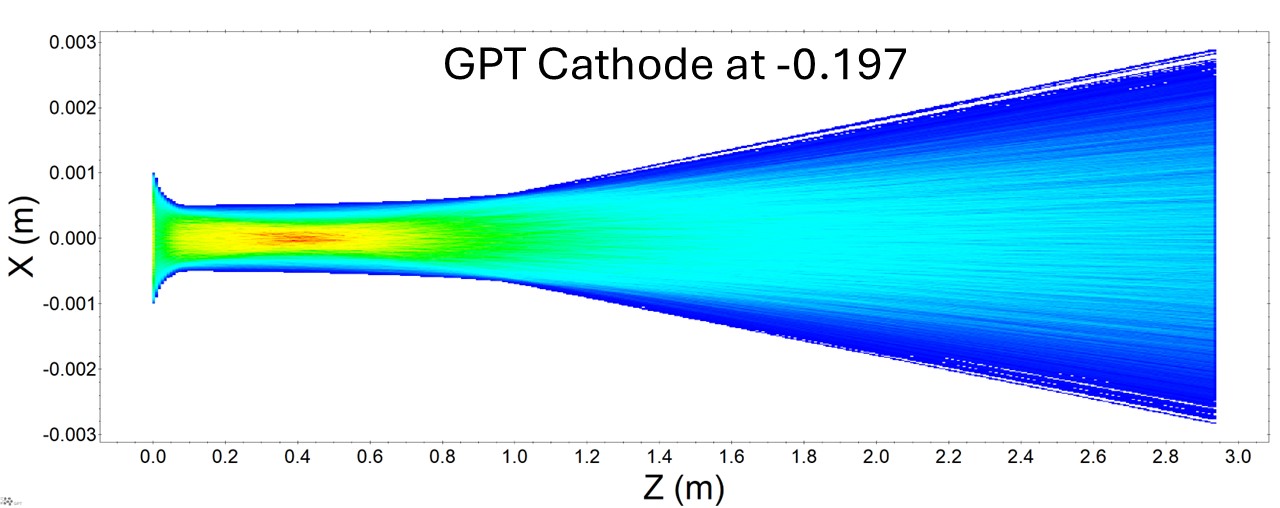}
	\caption{(a)GPT simulation with LEI cathode flush at 0 cm producing a large final bunch radius of 30 mm ; (b) GPT simulation with LEI cathode retracted to --0.197 cm producing a much smaller  bunch radius of 3 mm, at $z$~=~3m.}
	\label{GPT_x_z_0.197}
\end{figure}

For the same emission radius and launch phase, the retracted cathode produces a much smaller bunch radius of 3 mm, as compared to the un-retracted simulation with a bunch radius of 30 mm, both at $z$=~3~m. The retracted cathode simulation generates a beam waist approximately 0.4~m downstream from the cathode. In contrast, with the cathode at its nominal position, there is a virtual beam waist several cm upstream of the cathode, and the far-field divergence angle is approximately an order of magnitude larger. The somewhat peculiar shape of the bunch's edge in phase space is attributed to nonlinear components in the near-cathode radial rf fields; but the overall divergence is dominated by the intrinsic emittance of the beam, not the rf fields (regardless of linearity). The cathode retraction thus provides a phase space distribution which can be more accurately measured by a two-slit scan to yield an MTE.

\begin{figure}
	\includegraphics[width=7cm]{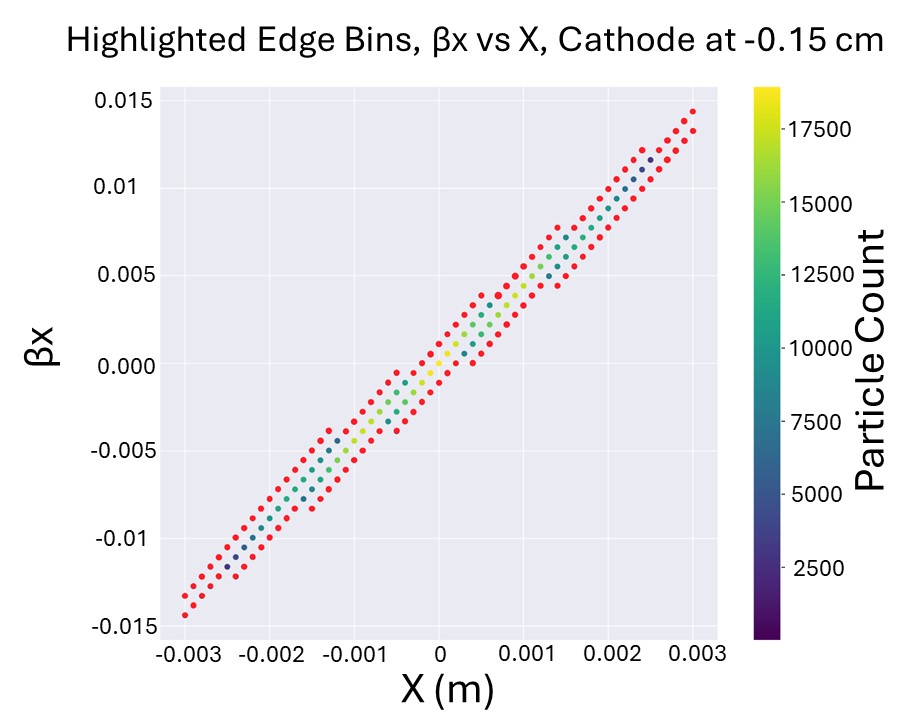}
	\caption{Edge bins of two-slit simulation with cathode at --0.15 cm 
 are marked in red. The total is 112 edge bins.}
	\label{Bx_0.15cm_Edge_Cells}
\end{figure}

\begin{figure}
	\includegraphics[width=7cm]{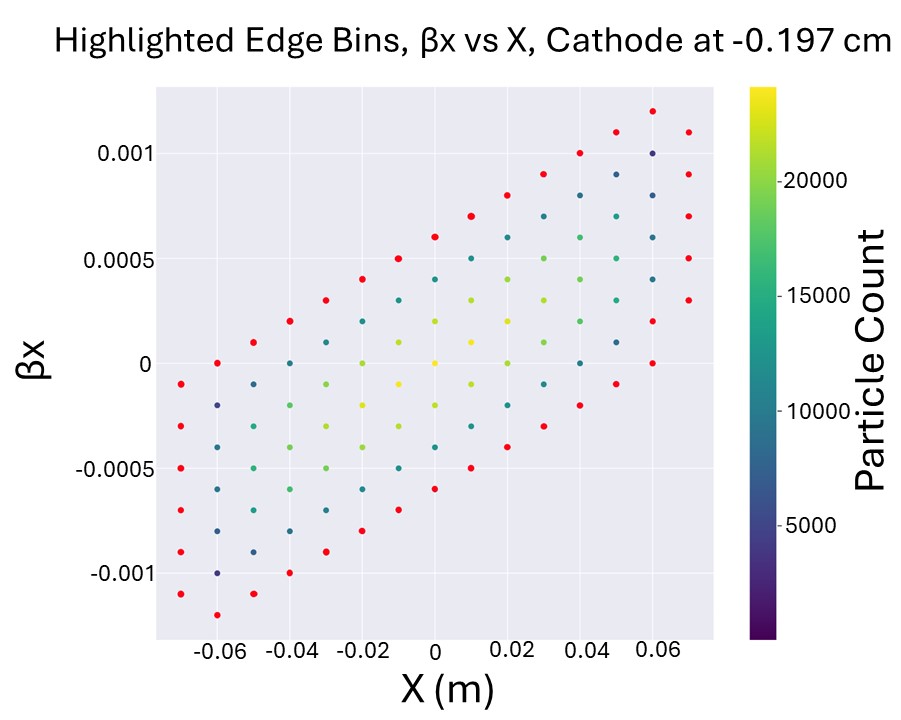}
	\caption{Edge bins of two-slit simulation with cathode at --0.197 cm 
 are marked in red. The total count is 38 edge bins.}
	\label{Bx_0.197cm_Edge_Cells}
\end{figure}

The difference in the number of edge bins between distributions from an ideally retracted vs. partly retracted cathode is drastic, as illustrated in Figs.\ref{Bx_0.15cm_Edge_Cells}~and~\ref{Bx_0.197cm_Edge_Cells}. There is approximately 3-fold difference in the number of edge bins indicating a similar drop in the measurement error. (We note that when the cathode is flush, the resulting strongly diverging distribution is effectively all contained in the edge bins.)  Together these effects -- fewer edge bins, and smaller extents in phase space -- allow to preform a measurement of the intrinsic emittance with lower systematic error.

\section{Discussion}\label{Discus}
\begin{figure}
	\includegraphics[width=7cm]{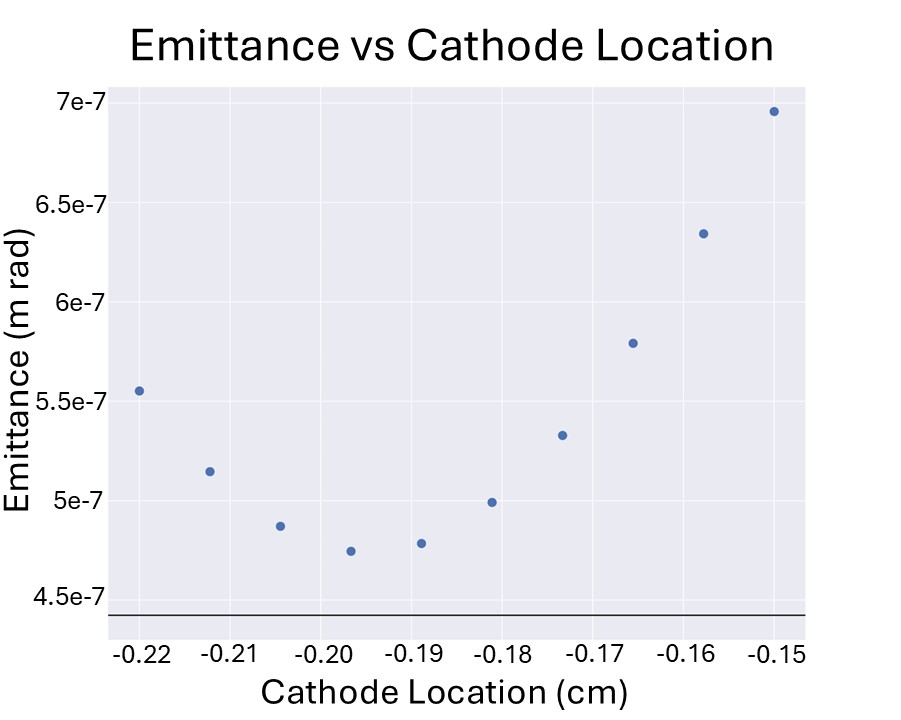}
	\caption{The simulated RMS emittance using the two-slit method for different cathode retraction location.}
	\label{Emittance_vs_Cathode_Location}
\end{figure}

\begin{figure}
	\includegraphics[width=7cm]{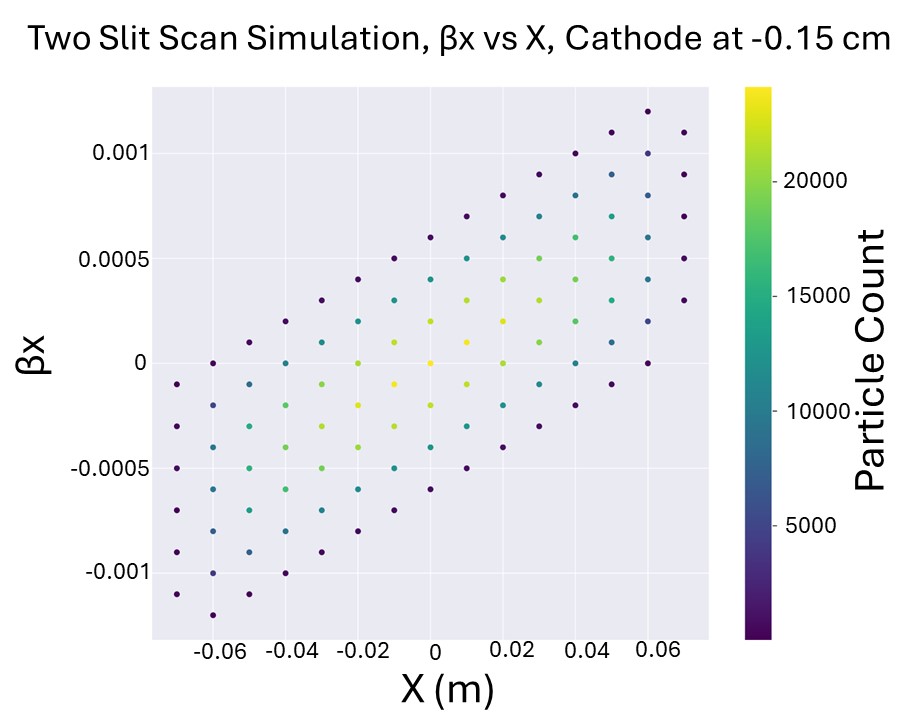}
	\caption{The simulated $x$-$x'$ phase space using the two slit method with the cathode positioned at --0.197 cm.}
	\label{Bx_0.197cm}
\end{figure}

\begin{figure}
	\includegraphics[width=7cm]{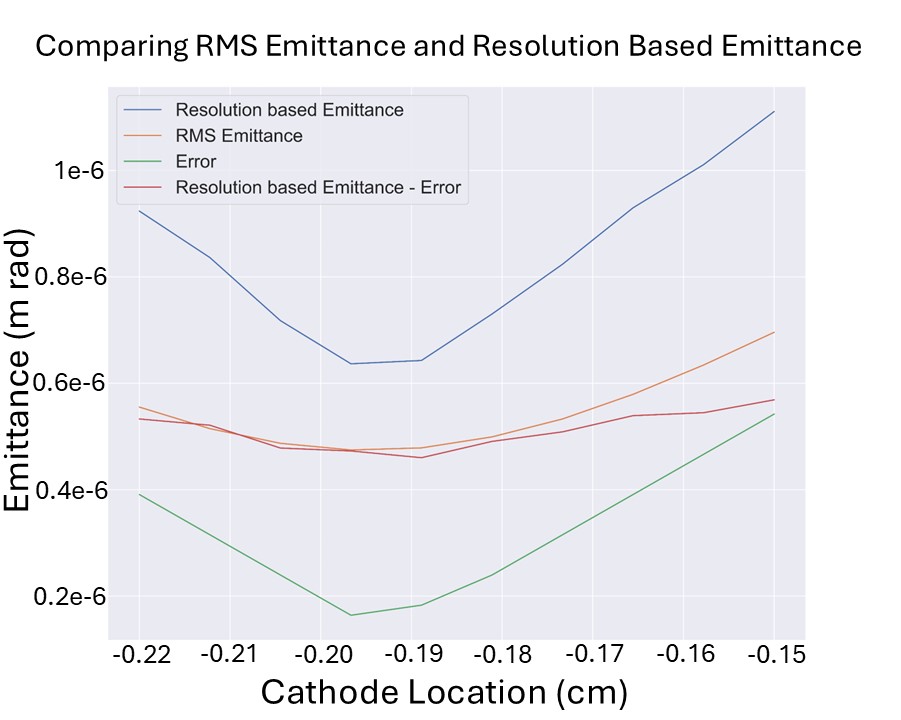}
	\caption{A comparison between the resolution based emittance and the simulated RMS emittance from the two-slit method at different cathode location.}
	\label{Resolution_based_Emittance}
\end{figure}

The results of the case study show that the minimal \textit{measured} emittance was found with the cathode retracted to --0.197cm, Fig.~\ref{Emittance_vs_Cathode_Location}. At this location, emittance was calculated as 0.475 $\mu$rad, corresponding to an MTE of 230 meV, 30 meV higher than would be expected solely from the cathode intrinsic emittance as marked by the black line in Fig.~\ref{Emittance_vs_Cathode_Location}. This suggests the cathode retraction method can provide a reasonable measure of cathode MTE, when the cathode installed in an operational high-brightness photoinjector. However, the measurement error of +15\% is still substantial. When the resulting phase space analysis is performed using the resolution-based emittance with consideration of the binning error as discussed above, in Fig.~\ref{Resolution_based_Emittance} there is nearly perfect agreement between the RMS emittance and the resolution based emittance when the binning error is subtracted. The calculation of the binning error also indicates that it is lowest at --0.197 cm due to the size of the bunch in phase space. Comparisons between Fig.~\ref{Phase_Space_0.197} and Fig.~\ref{Bx_0.197cm}, particle-based phase space and bin-based phase space respectively, show excellent agreement between the two, while tracking individual particles in GPT provides a more precise map of the phase space but is not feasible in a practical experimental system. 

The results highlight two particularly useful intertwined insights for a practical measurement: cathode retraction can be used to generate a beam in which the divergence is dominated by the cathode MTE; this is the same foundation upon which operation of instruments such as the Momentratron rely. This condition in turn, leads to a lower binning error in the two-slit measurement and, ultimately, to sub-20\% measurement errors for measuring the cathode MTE. In terms of practical implementation and experimental setup, the optimal cathode location to make an intrinsic emittance measurement is found simply via minimizing the spot size of the beam at the first slit location. This provides a fast way to identify the optimal cathode position for the measurement. Second, by effectively compensating for rf defocusing within the remainder of the gun, cathode retraction allows the MTE to become the dominant term impacting the measurement process. Together theses effects are sufficient for measuring the MTE of the cathode within a reasonably small margin of error.  Analysis of the systematic error contributions provide understanding of why these effects decrease the measurement error. Indeed, calculating the resolution based emittance and binning error provides not only a useful double check of an RMS emittance measurement but explains why an emittance measurement will have lower error with a smaller, more slowly diverging beam than one where the bunch is strongly diverging, e.g. due to rf fields. By analyzing the binning error equation, Eq. \eqref{Error}, we observe that bunches with larger extents in phase space, all else equal, will have larger measurement errors than bunches with the same phase-space area, but smaller extents in phase space. Understanding the error contribution that different phase space shapes will contribute provides a critical tool in determining the utility of a two-slit measurement.

\section{Conclusion and Outlook}\label{Conc}
This work introduces an improved two-slit emittance measurement methodology that combines several techniques to measure the intrinsic emittance of a beam, and thus the MTE of the photocathode from which it was emitted. These techniques have been simulated to show their effectiveness at isolating the intrinsic emittance. As is often the case the approach to simulating a system is different from using that system in practice. The limitations for a two-slit measurement can be broadly categorized as duration based and error based. The discussed case study provided ways to address both of these limitations so that a practical system can be proposed.

The duration limitation can be addressed in part by utilizing a simple approach to determine the preferred cathode position for an emittance measurement, as only the diameter of the beam needs to be measured to set the cathode position.  This eliminates the need to make slit-scan measurements at multiple cathode locations.  Careful design of the slit system, e.g. determining the optimal number of bins, will help to minimize the time required to make a single scan while maintaining the desired resolution. While a thorough analysis of minimizing measurement time is beyond the scope of this paper, we note several possible directions. Increased bunch charge can decrease measurement time (e.g. accumulating the same statistics, in terms of charge per bin etc., more quickly) albeit at the expense of increased space-charge contributions to the beam emittance; this can be explored further in simulation.  Substituting fast beam steering (via magnetic or electrostatic deflectors) for physical slit motion \cite{lewellen_electrostatic_2018, bazarov_benchmarking_2008} could potentially significantly decrease the measurement time; the steering fields do not perturb the phase space and thus does not corrupt the measurement.  Incorporation of intelligent algorithms into the measurement process, e.g. to identify likely regions of no current density, can reduce the number of total locations to be sampled in phase space.

The error limitation of two-slit measurements cannot be avoided as binning of the particles will occur based of the size of the slits used. However, as is shown here, it is possible to generate phase space distributions where the binning error is minimized and an RMS measurement of emittance is not strongly affected by the binning error. The phase space effects shown here are attributed to the dynamics at play from cathode retraction and work ideally for measuring the intrinsic emittance. In general, the resolution-based emittance and binning error estimates may provide useful corrections to the RMS emittance calculated from the measured phase-space map.    

\section{Acknowledgments}\label{Acknow}
The work by Benjamin Sims was supported by the U.S. Department of Energy Office of Science, High Energy Physics under Cooperative Agreement Award No. DE-SC0018362. The work by Sergey Baryshev was supported by the U.S. Department of Energy, Office of Science, Office of High Energy Physics under Award No. DE-SC0020429. The work was supported by the U.S. Department of Energy Office of Science, High Energy Physics under Cooperative Agreement Award No. DE-SC0018362, and the US Department of Energy under Contract DE-AC02-76SF00515.

\bibliography{references}

\end{document}